\pdfoutput=1

\documentclass[final, numberedheadings, sort&compress, a4paper]{aipproc}

\usepackage{amsmath,amsthm,amsfonts}
\usepackage{amssymb}
\layoutstyle{6x9}

\usepackage{graphicx}
\usepackage{epstopdf} 

\allowdisplaybreaks[4]
\begin{document}

\title{Origins of the Combinatorial Basis of Entropy}

\classification{
02.50.Cw, 
02.50.Tt, 
05.20.-y, 
05.70.-a, 
05.90.+m, 
89.20.-a, 
89.70.+c 
}

\author{Robert K. Niven}{
  address={(1) School of Aerospace, Civil and Mechanical Engineering, The University of New South Wales at ADFA, 
Canberra, ACT, 2600, Australia. Email: r.niven@adfa.edu.au}
,altaddress={(2) Niels Bohr Institute, University of Copenhagen, Copenhagen \O, Denmark.} }

\begin{abstract}
\noindent
The combinatorial basis of entropy, given by Boltzmann, can be written $H =  N^{-1} \ln \mathbb{W}$, where $H$ is the dimensionless entropy, $N$ is the number of entities and $\mathbb{W}$ is number of ways in which a given realization of a system can occur (its statistical weight). This can be broadened to give generalized combinatorial (or probabilistic) definitions of entropy and cross-entropy: $H=\kappa (\phi(\mathbb{W}) +C)$ and $D=-\kappa (\phi(\mathbb{P}) +C)$, where $\mathbb{P}$ is the probability of a given realization, $\phi$ is a convenient transformation function, $\kappa$ is a scaling parameter and $C$ an arbitrary constant. If $\mathbb{W}$ or $\mathbb{P}$ satisfy the multinomial weight or distribution, then using $\phi(\cdot)=\ln(\cdot)$ and $\kappa=N^{-1}$, $H$ and $D$ asymptotically converge to the Shannon and Kullback-Leibler functions. In general, however, $\mathbb{W}$ or $\mathbb{P}$ need not be multinomial, nor may they approach an asymptotic limit.  In such cases, the entropy or cross-entropy function can be {\it defined} so that its extremization (``MaxEnt'' or ``MinXEnt"), subject to the constraints, gives the ``most probable'' (``MaxProb'') realization of the system. This gives a probabilistic basis for MaxEnt and MinXEnt, independent of any information-theoretic justif\-ication. 

This work examines the origins of the governing distribution $\mathbb{P}$.  These include:\ (a) frequentist-like models; (b) symmetry models; (c) prior MinXEnt models; (d) Kapur-Kesavan inverse models; and (e) game theoretic models.  The combinatorial definition and MaxProb are consistent with these different approaches, and the notion of probabilistic inference, yet offer greater utility than traditional MaxEnt / MinXEnt based on the Shannon and Kullback-Leibler functions.
\end{abstract}

\keywords{MaxEnt; MaxProb; Boltzmann principle; combinatorial; probabilistic inference}
\maketitle

\section{\label{sect:intro}Introduction} 
%
Fifty years ago, Jaynes \cite{Jaynes_1957} gave the maximum entropy method (MaxEnt), based on the Shannon entropy \cite{Shannon_1948}:
\begin{equation}
H_{Sh} = - \sum\limits_{i = 1}^s {p_i \ln p_i } 
\label{eq:Shannon}
\end{equation}
where $p_i$ is the (posterior) probability of occurrence of the $i$th distinguishable state within a system, from $s$ such states.
In the MaxEnt method, one maximizes the Shannon entropy of a system, subject to its constraints, to determine the ``least informative'' or ``maximally noncommittal'' probability distribution representing the system.  From its inception, MaxEnt was advanced as a generic method of inference for the solution of indeterminate problems of all kinds, underpinned by information theory, not merely as an extension of mechanics \cite{Jaynes_1957, Jaynes_1963, Jaynes_1968, Jaynes_1978, Jaynes_2003}. MaxEnt was later extended into the maximum relative entropy, minimum divergence or minimum cross-entropy method (MinXEnt), involving extremization of the Kullback-Leibler measure \cite{Kullback_L_1951, Kullback_1959}: \pagebreak 
\begin{equation}
D_{KL} = \sum\limits_{i = 1}^s {p_i \ln \frac{{p_i }}{{q_i }}} 
\label{eq:KL}
\end{equation}
which allows for unequal prior probabilities $q_i$. Since that time, MinXEnt and its subsidiary MaxEnt have been successfully applied to the analysis of a vast number of phenomena, throughout most fields of human study \citep[e.g.][]{Jaynes_2003, Levine_T_1978, Kapur_1989b, Kapur_K_1992}, and can rightly be regarded as one of the most important of all human discoveries.

It must be emphasised, however, that the cross-entropy and entropy concepts which underpin MinXEnt and MaxEnt are themselves subject to many different philosophical interpretations. Dominant explanations include the axiomatic basis outlined by Shannon \cite{Shannon_1948}, and the information-theoretic (``bits'' of information) and coding basis, recognized by Szilard \cite{Szilard_1929} and Shannon \cite{Shannon_1948} \citep[c.f.][]{Shore_J_1980}. These bases led Jaynes, in particular, to consider the Shannon and Kullback-Leibler functions to be the only logically consistent measures of uncertainty, and thus the only ones suitable for analysis. This view has been challenged by many researchers, on the grounds that the above two measures are too narrowly defined and/or inapplicable to many situations. For example, over the past 85 years, many alternative entropy and divergence functions have been introduced \citep[e.g.][]{Fisher_1922, Fisher_1925, Bose_1924, Einstein_1924, Einstein_1925, Fermi_1926, Dirac_1926, Renyi_1961, Sharma_M_1975, Sharma_M_1977, Tsallis_1988, Tsallis_2001, Kaniadakis_2001, Kaniadakis_2002, Beck_C_2003, Niven_2005, Niven_2006}; in most cases, these are incompatible with the Shannon and Kullback-Leibler functions, but have proved {\it useful} for the analysis of specific classes of systems. Can such measures be explained by some broader philosophical framework? How should we choose the ``correct'' cross-entropy or entropy function for a given problem?  The fact that such questions remain unanswered indicates the need for a unifying philosophical framework, which encompasses (and {\it explains}) such alternative entropy measures and their connections to information theory.

This study examines one such framework:\ the combinatorial (or probabilistic) basis of entropy, first given 130 years ago by Boltzmann \cite{Boltzmann_1877} and subsequently promoted by Planck \cite{Planck_1901}. This involves the maximization of a governing probability distribution $\mathbb{P}$ or weight $\mathbb{W}$ of a system; this can be viewed as a generalized principle of probabilistic inference, aptly described by Vincze and Grendar \& Grendar as the maximum probability (``MaxProb'') principle \cite{Vincze_1972, Grendar_G_2001}. It also leads to generalized definitions of cross-entropy and entropy, based purely on probability theory \cite{Niven_CIT}.  In this study, specific attention is paid to the origins of the governing distribution $\mathbb{P}$, including (a) frequentist-like models (e.g.\ ball-in-box or urn models); (b) symmetry models; (c) prior MinXEnt models; (d) Kapur-Kesavan inverse models; and (e) game theoretic models.  It is shown that the combinatorial basis is consistent with these different approaches, but is more soundly based and offers greater utility than traditional MaxEnt / MinXEnt based on the Shannon and Kullback-Leibler functions.
\section{\label{sect:comb}The Combinatorial Basis} 
%
\begin{figure}[t]
\setlength{\unitlength}{0.6pt}
  \begin{picture}(450,240)
   \put(0,0){\includegraphics[width=100mm]{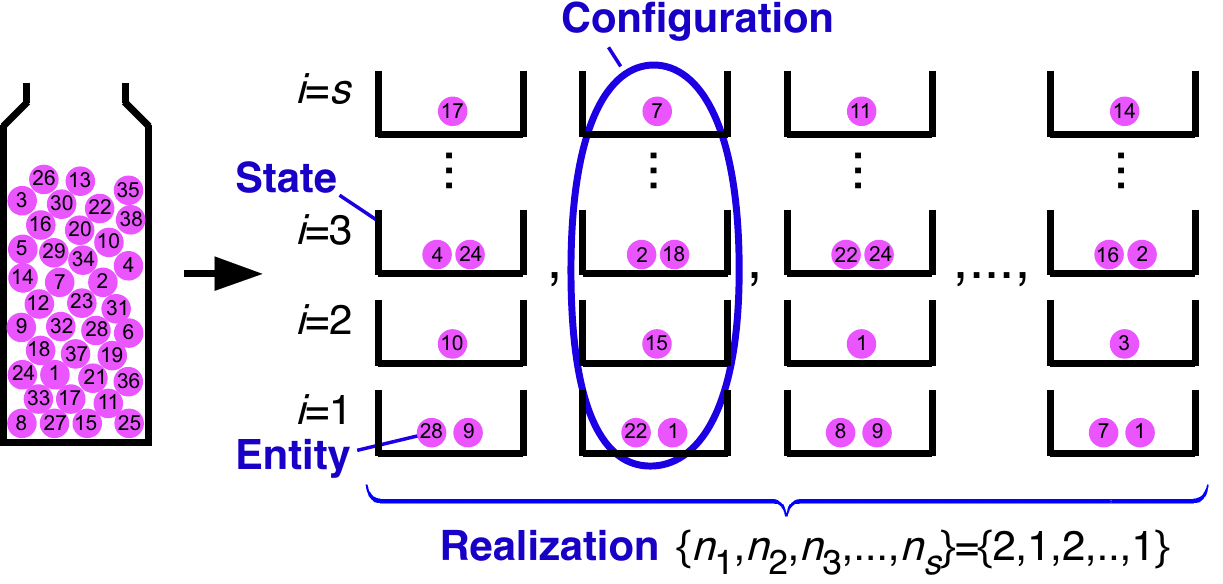} }
  \end{picture}
\caption{Definition of terms used in the combinatorial basis of entropy and cross-entropy.}
\label{fig:defs}
\end{figure}

Owing to a tremendous confusion in terminology - especially amongst physicists - it is first necessary to rigorously define several important terms \citep[c.f.][]{Niven_CIT}.  An {\it entity} is here taken to be a discrete particle, object or agent within a system, which acts separately but not necessarily independently of the other entities present.  A {\it system} is a collection of entities with a defined boundary, subject to various constraints, which may or may not be open to the exchange of specified entities or substances with an external environment.  The entity therefore constitutes the unit of analysis of a system.  

Now consider a simple ``ball-in-box'' model of a system, shown in Figure \ref{fig:defs}, in which $N$ {distinguishable} entities (balls) are allocated to $s$ {distinguishable} non-degenerate states (boxes).  As shown:
\begin{list}{$\bullet$}{\topsep 2pt \itemsep 2pt \parsep 0pt \leftmargin 8pt \rightmargin 0pt \listparindent 0pt
\itemindent 0pt}
\item A {\it state} refers to each different category or element of system (e.g.\ energy levels, sides of a die or alphabetic symbols). The states are therefore properties of, or associated with, each individual entity in the system.  
\item A {\it configuration} is a distinguishable permutation or pattern of entities amongst the states of a system (a {\it complexion}, {\it microstate} or {\it sequence}). A configuration is therefore a property of the system as a whole.
\item A {\it realization} is each aggregated arrangement of entities amongst the states of a system, as specified by some rule, for example by the number of entities in each state (a {\it macrostate}, {\it outcome} or {\it type}). In general, a realization will constitute a set of configurations, since several configurations could give the same realization (see Figure \ref{fig:defs}). 
\end {list}
There is such confusion in and sloppy usage of the terms {\it state}, {\it microstate} and {\it macrostate} - severely impairing understanding - that the last two terms should be avoided. In the following, the states are indexed $i=1,...,s$ (which may be multivariate); $n_i$ denotes the number of entities in the $i$th state; $q_i$ and $p_i=n_i/N$ respectively denote the prior and posterior probabilities of a entity being in the $i$th state; and each realization\footnote{A realization can only be denoted $\{p_i\}$ in the asymptotic limits $N \to \infty$ and $n_i \to \infty, \forall i$, since $\{p_i\}$ discards information about the value of $N$.} is denoted $\{n_i\}$.  Notwithstanding other philosophical differences with Jaynes, the ``subjective Bayesian'' definition of probabilities, as assignments based on what we know, is adopted here \cite{Jaynes_1957,Jaynes_2003}. 

%
%
For the analysis of probabilistic systems, it is possible to delineate a principle which stands out from all others: the {\it maximum probability} (``MaxProb'') principle \cite{Boltzmann_1877, Planck_1901, Vincze_1972, Grendar_G_2001, Niven_CIT}. This can be stated as: 
\begin{list}{~}{\topsep 3pt \itemsep 0pt \parsep 0pt \leftmargin 8pt \rightmargin 0pt \listparindent 0pt
\itemindent 0pt}
\item {\it ``A system can be represented by its realization of highest probability.''}
\end{list}
This seemingly trivial statement provides a powerful principle for {\it probabilistic inference}, which is independent of any information-theoretic considerations. This is critical, since in any contradiction between information theory and probability theory - for example, between the distributions inferred by each approach - probability theory must triumph. Like MinXEnt or MaxEnt based on the Kullback-Leibler or Shannon measures, MaxProb is a method of inference (inductive reasoning), which does not give certainty in its predictions.  Unlike them, however, MaxProb is founded solely on probability theory.  Indeed, MaxProb does not depend upon any asymptotic limits (a feature of the ``frequentist'' definition of probability, in which probabilities must correspond to measurable frequencies \cite{Jaynes_1957, Jaynes_2003}); it can therefore be applied to systems containing finite numbers of entities \cite{Niven_2005, Niven_2006}.  

Allied to MaxProb is a generalized form of the second law of thermodynamics:
\begin{list}{~}{\topsep 3pt \itemsep 0pt \parsep 0pt \leftmargin 8pt \rightmargin 0pt \listparindent 0pt
\itemindent 0pt}
\item {\it ``A system will tend towards its most probable realization.''}
\end{list}
This provides a purely probabilistic rationale for use of the MaxProb principle, independent of thermodynamics.  In effect, if we adopt MaxProb as a principle for probabilistic inference, the above statement is its corresponding ergodic principle, which (on average) explains its success.  Of course - as expressed by Jaynes \cite{Jaynes_1957} - the concept of ergodicity is not needed for the purpose of inference, since in the absence of other information, we are fully justified in conducting inference without it.  

\begin{figure}[t]
\setlength{\unitlength}{0.6pt}
  \begin{picture}(350,330)
   \put(0,0){\includegraphics[width=75mm]{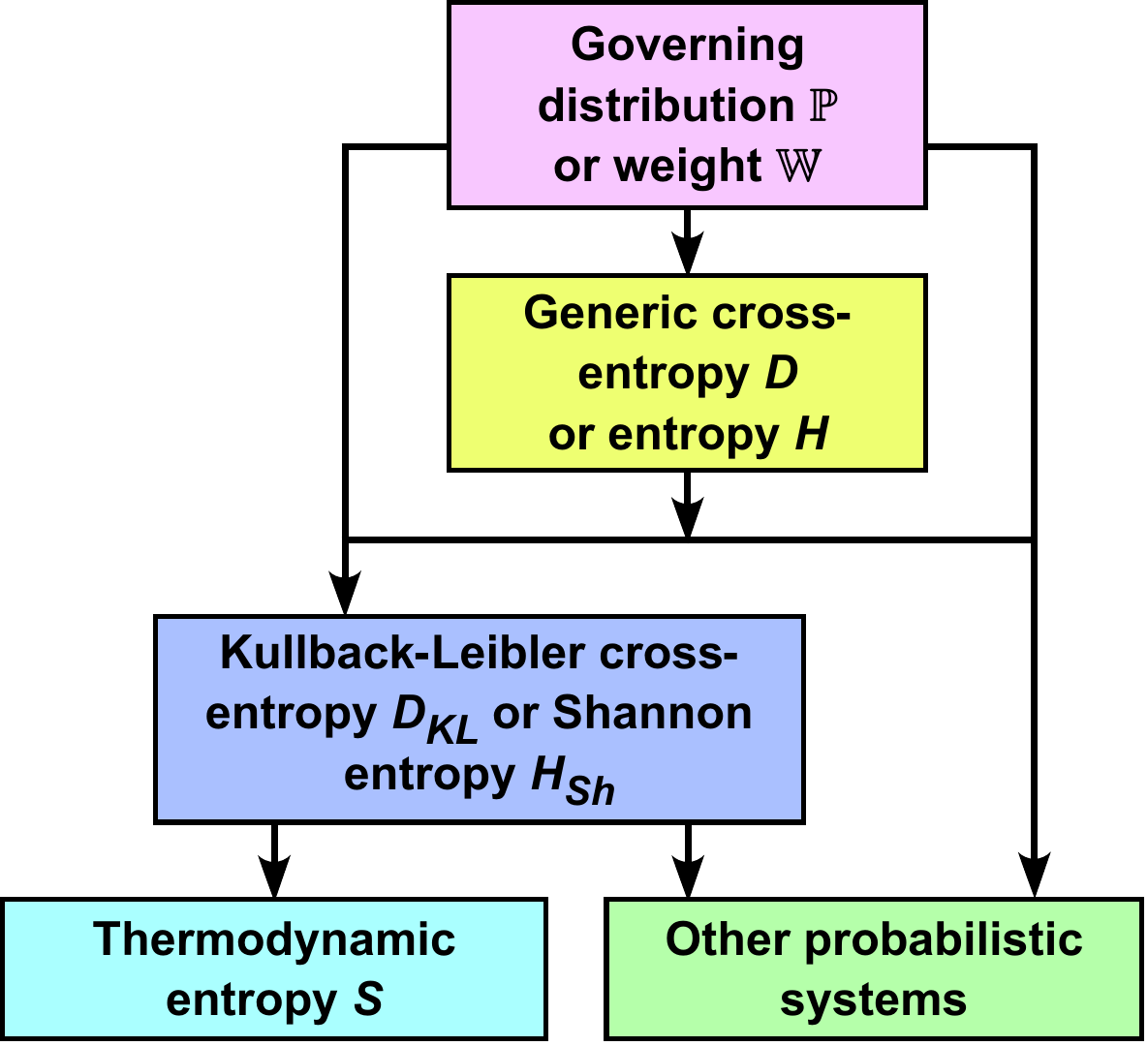} }
  \end{picture}
\caption{Schematic flowchart of the combinatorial basis of entropy and cross-entropy.}
\label{fig:flowchart}
\end{figure}

%
%
The MaxProb principle also leads to the {\it combinatorial definition} of entropy, first given by Boltzmann \cite{Boltzmann_1877} and Planck \cite{Planck_1901}.  This can be written as: 
\begin{equation}
H ={N}^{-1} \ln \mathbb{W},
\label{eq:Boltzmann}
\end{equation}
where $\mathbb{W}$ is the number of ways in which a given realization can occur, referred to as its statistical weight.  Maximization of the entropy $H$ of a system, subject to its constraints, therefore selects the realization of highest weight $\mathbb{W}$ (the logarithmic function being a monotonic transformation, which does not alter the position of the extremum).  Eq.\ \eqref{eq:Boltzmann} can be extended to give generalized combinatorial (or probabilistic) definitions of cross-entropy and entropy \cite{Niven_CIT}: 
\begin{gather}
D=-\kappa (\phi(\mathbb{P}) +C), \qquad  H=\kappa (\phi(\mathbb{W}) +C),
\label{eq:gencombEnt}
\end{gather}
where $\mathbb{P}=P(\{n_i\} | \{q_i\},N,s,I) $ is the probability of a given realization, subject to the prior probabilities $\{q_i\}$, number of entities $N$, number of states $s$ and background information  $I$; $\phi$ is a convenient monotonic transformation function; $\kappa$ is a scaling parameter; and $C$ is an arbitrary constant. This perspective is summarised in Figure \ref{fig:flowchart}. If $\mathbb{P}$ or $\mathbb{W}$ satisfy the multinomial distribution or weight:
\begin{gather}
\mathbb{P} = N! \prod\limits_{i = 1}^s \frac{q_i ^{n_i } } {n_i !} ,
\qquad \mathbb{W} = N! \prod\limits_{i = 1}^s \frac{1 }{n_i !}, 
\label{eq:multinomialWt}
\end{gather}
then by taking $\phi(\cdot)=\ln(\cdot)$, $\kappa=N^{-1}$ and the asymptotic limits $N \to \infty$ and $n_i \to \infty, \forall i$ (the ``Stirling approximation''), $D$ and $H$ converge respectively to the Kullback-Leibler and Shannon functions \eqref{eq:Shannon}-\eqref{eq:KL} \cite{Vincze_1972, Grendar_G_2001}. This provides a (well-known) justification for these functions, and their corresponding MinXEnt and MaxEnt principles, as a special case, independently of the arguments used in information theory.

In general, however, $\mathbb{P}$ or $\mathbb{W}$ need not be multinomial, nor may they approach an asymptotic limiting form.  In such cases, extremization of the cross-entropy or entropy {defined} by \eqref{eq:gencombEnt}, subject to the constraints, gives the most probable (MaxProb) realization of the system (in the non-asymptotic case, due to the effect of quantization, extremization gives an ``attractor'' distribution which lies close to but not necessarily equal to the MaxProb realization \cite{Niven_CIT}). In consequence, the combinatorial definitions \eqref{eq:gencombEnt} remain consistent with the rules of probability theory, whilst inference using the Kullback-Leibler or Shannon measures may lead to inconsistencies. The combinatorial (or probabilistic) definitions are therefore more broadly applicable than those derived from information-theoretic considerations. 

In the foregoing discussion, the astute reader will notice that there may be many different ways to classify the entities and states of a system, and hence to identify its configurations; and many different ways to group the configurations into realizations.  We are therefore led to the ``subjective'' (or ``observer-dependent'') view of the entropy and cross-entropy concepts, a sentiment vocally defended by Jaynes \cite{Jaynes_1957}.  This was succinctly expressed by Tseng and Caticha \cite{Tseng_C_2002}:
\begin{list}{~}{\topsep 6pt \itemsep 0pt \parsep 0pt \leftmargin 8pt \rightmargin 0pt \listparindent 0pt
\itemindent 0pt}
\item {\it ``Entropy is not a property of a system $\dots$ [it] is a property of our description of a system.''}
\end{list}
The fact that the thermodynamic entropy $S$ is always defined in the same manner, allowing thermodynamicists to make consistent calculations, should not fool the reader into believing that the entropy concept is ``objective''\footnote{It is important that the symbol $S$ be devoted exclusively to the thermodynamic entropy, since it is a special case of - but is distinct from - the dimensionless Shannon entropy \eqref{eq:Shannon}.}.  
\pagebreak

\section{\label{sect:origins}Origins of the Governing Distribution} 
%
We now consider the origins of the governing distribution $\mathbb{P}$ or weight $\mathbb{W}$ used in the combinatorial formulation. The problem of justifying a cross-entropy or entropy function is now replaced by a deeper problem, of how to justify its governing distribution. The ubiquity of the Kullback-Leibler and Shannon measures, in many circumstances, therefore leads to the question: why the multinomial distribution? This question, and the choice of $\mathbb{P}$, is examined from five different perspectives. 

\vspace{6pt}
\noindent {\bf (a) Frequentist-like Models}

In this approach, one simply asserts a governing distribution $\mathbb{P}$ or weight $\mathbb{W}$ as a probabilistic model of the system under consideration. One may have strong grounds, based on prior knowledge of a problem, for such an assertion; in any case, we should have no ``fear of failure'' of this method (in Jaynes' words), since if the model gives unsuccessful predictions, we have learnt that it is incorrect.  Many such models are available from classical probability theory, for example the {\it ball-in-box} models of the type represented in Figure \ref{fig:defs}.  Since these arose from frequentist studies, they can be termed ``frequentist-like'' models, although used here for the purpose of inference.

In the case discussed previously, in which distinguishable balls are allocated to distinguishable boxes in accordance with a set of constant prior probabilities (see Figure \ref{fig:defs}), one obtains the multinomial distribution \eqref{eq:multinomialWt}, and hence the Kullback-Leibler cross-entropy and Shannon entropy functions in the Stirling limits.  However, different assumptions lead to different model distributions. If the asymptotic limits are not applied, then from \eqref{eq:multinomialWt}, one obtains a non-asymptotic cross-entropy function \citep[c.f.][]{Niven_2005}:
\begin{align}
\begin{split}
  - D_{KL}^x  &= {N}^{-1} \ln \mathbb{P}  = {N}^{-1} \bigl\{ \ln N! + \sum\limits_{i = 1}^s {n_i \ln q_i}  -  \sum\limits_{i = 1}^s {\ln n_i ! } \bigr\} \\ 
    &= \sum\limits_{i = 1}^s {\bigl\{ { {p_i }{N}^{-1} \ln N! + p_i \ln q_i  - {N}^{-1} \ln [(p_i N)!]} \bigr\}}  \\ 
 \end{split}
\end{align}
This is applicable to systems with finite (small) $N$.  Minimisation of $D_{KL}^x$, subject to the usual constraints $\sum\nolimits_{i=1}^s n_i=N$ and $\sum\nolimits_{i=1}^s n_i f_{ri}=N \langle f_r \rangle$, for $r=1,...,R$, where $f_{ri}$ is the $r$th function of each state $i$ and $\langle f_r \rangle$ is its mathematical expectation, gives the ``most probable'' distribution \citep[c.f.][]{Niven_2005}: 
\begin{equation}
p_i ^{\#} = {N}^{-1} \Bigl[ \psi ^{ - 1} \bigl( {N}^{-1} \ln N! + \ln q_i  - \lambda _0  - \sum\limits_{r = 1}^R {\lambda _r f_{ri} }  \bigr) - 1 \Bigr]
\label{eq:exact}
\end{equation}
where $\psi ^{ - 1}(\cdot)$ is the inverse digamma function. Eq.\ \eqref{eq:exact} can be viewed as the ``attractor'' for systems with finite $N$, which differs from the attractor given by traditional MinXEnt. 

If the states are considered to contain $g_i$ distinguishable, degenerate sub-states within each distinguishable state $i$, then three cases have been examined historically: (i) distinguishable entities; (ii) indistinguishable entities; and (iii) indistinguishable entities, with a maximum of one entity in each state. The resulting distributions were given by Brillouin \cite{Brillouin_1927, Brillouin_1930} as, respectively: \pagebreak
\begin{align}
\mathbb{P}_{MB} &= \frac{N!}{G^N} \prod\limits_{i = 1}^s {\frac{{g_i ^{n_i } }}{{n_i !}}} ,
\label{eqPu_MB} \\
\mathbb{P}_{BE} &= \frac{{N!(G - 1)!}}{{(G + N - 1)!}}\prod\limits_{i = 1}^s {\frac{{(g_i  + n_i  - 1)!}}{{n_i !(g_i  - 1)!}}},
\label{eqPu_BE} \\
\mathbb{P}_{FD} &= \frac{{N!(G - N)!}}{{G!}}{\rm{ }}\prod\limits_{i = 1}^s {\frac{{g_i !}}{{n_i !(g_i  - n_i )!}} }.
\label{eqPu_FD} 
\end{align}
where $G= \sum\nolimits_{i = 1}^s {g_i}$ is the total degeneracy.  The truncated weights and entropy functions corresponding to these distributions, referred to respectively as the Maxwell-Boltzmann, Bose-Einstein and Fermi-Dirac distributions respectively \citep[e.g][]{Bose_1924, Einstein_1924, Einstein_1925, Fermi_1926, Dirac_1926, Brillouin_1927, Brillouin_1930, Tolman_1938, Brillouin_1951b, Davidson_1962}, played an important role in the development of quantum theory.  In the non-asymptotic case, the resulting entropy functions appear to have profound information-theoretic consequences \cite{Niven_2005, Niven_2006}. 

Recently, a quite different ball-in-box model was considered, in which distinguishable entities are allocated to indistinguishable, equally degenerate states.  The statistical weight of each realization $\{n_i\}$ can be expressed as \cite{Niven_Catania}:
\begin{align}
\mathbb{W}_{D:I(g)}   
= \frac {N!} { \Bigl( \prod\limits_{i=1}^k  {n_i!} \Bigr) \Bigl( \prod\limits_{j=1}^N {r_j!} \Bigr) }  \prod\limits_{i=1}^k \hspace{4pt}  \sum\limits_{\gamma=1}^{\min(g,n_i)} \Bigl \{  \begin{matrix} n_i \\  \gamma  \\  \end{matrix} \Bigr \}
\label{eq:ind_state_wt}
\end{align}
where there are $k$ non-empty states amongst the $s$ states; $g$ is the degeneracy of each state; ${\bigl\{  \begin{smallmatrix}    {n_i }  \\  \gamma  \\ \end{smallmatrix} \bigr\}}$ is a Stirling number of the second kind; and $r_j$ is the number of occurrences of integer $j$ in the set $\{n_i\}$. The combinatorial entropy corresponding to \eqref{eq:ind_state_wt}, $H_{D:I(g)}=N^{-1} \ln \mathbb{W}_{D:I(g)}$, does not appear to have a straightforward asymptotic form, except in the non-degenerate case $g=1$ with $k=s$, when it reduces to the Shannon entropy.

Closely related to but distinct from ball-in-box models are {\it urn models}, in which a container (urn) is set up with a total of $M$ balls, made up of $m_i$ balls of each color $i$. Balls are then drawn from the urn in accordance with some sampling scheme, recorded and returned to the urn (or the urn modified in some way), and the sampling repeated \citep[c.f.][]{Jensen_1985, Berg_1988}. The asymptotic limits of an infinitely large urn ($M \to \infty$ and $m_i \to \infty, \forall i$), and an infinitely large (smaller) sample ($N \to \infty$ and $n_i \to \infty, \forall i$), are usually applied. Although quite different to the ball-in-box model of Figure \ref{fig:defs}, an urn model with simple replacement also yields the multinomial distribution \cite{Jensen_1985, Berg_1988}. Urn models involving the drawing of balls without replacement, or double replacement, lead respectively to the Fermi-Dirac and Bose-Einstein distributions \cite{Jensen_1985}. Urn models also readily permit the construction of systems in which the prior probabilities are not independently and identical distributed (non-{\it iid} sampling): e.g.\ the P\'olya distribution, in which after every draw, the ball is returned, and $c$ balls of the same color are also added \cite{EP, Steyn, JK, Grendar_N_Polya}:
\begin{equation}
\mathbb{P}_{Polya} = \frac{N!}{\prod\limits_{i=1}^s n_i!} 
\prod\limits_{i=1}^s \frac{m_i (m_i+c) \dots (m_i+(n_i-1)c)} {M(M+c) \dots (M+(N-1)c)},
\label{eq_Polya_distrib}
\end{equation}  
Substituting the {\it initial} prior probabilities $q_i=m_i/M$ and parameter $\beta = N/M$, this gives analytic cross-entropy measures in the non-asymptotic and asymptotic cases \cite{Grendar_N_Polya}. The resulting ``most probable'' distribution is intermediate between the Bose-Einstein and Fermi-Dirac distributions, with physical applications. 


\vspace{6pt}
\noindent {\bf (b) Symmetry-Based Arguments}

One may also choose a governing distribution on the basis of symmetry arguments (related to the ``principle of insufficient reason''). For a system made up of tosses of a coin, it is rational to consider the sampling to follow the binomial distribution, with equal prior probabilities of $\frac{1}{2}$ for each face, due to the symmetry of the states (there being no information to suggest that one state should be preferred). Alternatively, as suggested by David Blower at MaxEnt07, one can obtain a binomial distribution by the symmetry of all possible models in the model space (assigning a uniform prior to the models, over the entire spectrum from an all-head to an all-tail model, there being no reason to prefer any model). Applied to systems with more than two states, either argument leads to the multinomial distribution. In this respect, the multinomial distribution plays a role somewhat analogous to a central limit theorem (a ``central model theorem''), a point which deserves greater mathematical attention; this may be the reason for the ubiquity of the Kullback-Leibler and Shannon measures. Without symmetry, however, the argument breaks down, and one must adopt some other method to identify the governing distribution. 

\vspace{6pt}
\noindent {\bf (c) Prior MinXEnt Models}

A third origin of the governing distribution $\mathbb{P}$ is as a result of the application of MinXEnt at a higher level, for example to the set of systems within which the actual system resides. For example, the multinomial distribution can be obtained by MinXEnt based on the Kullback-Leibler cross-entropy, subject to a multinomial prior and mean constraints on each variate \cite{Kapur_K_1992}. This might then be imposed as a lower-level governing distribution. One can in fact envisage a hierarchy of governing and ``most probable'' distributions, at different levels of description. In a complex system, in which there is bidirectional feedback, the result will be a mosaic of interconnected probabilistic models (with thanks to the discussion by Tony Bell at MaxEnt07).

\vspace{6pt}
\noindent {\bf (d) Kapur-Kesavan Inverse Models}

The governing distribution $\mathbb{P}$ can also be obtained by extension of the arguments of Kapur and Kesavan \cite{Kapur_K_1992}, in which one works backwards from an observed probability distribution $\{p_i^*\}$, prior distribution $\{q_i\}$ (if available) and any constraints, to derive the measure of cross-entropy or entropy applicable to a system. By unravelling of the asymptotic limits, this could (at least in principle) be extended to determine the governing distribution of the system.  This avenue of research has not been examined in detail, and deserves greater attention.

\vspace{6pt}
\noindent {\bf (e) Game Theoretic Models}

The final method considered here is to derive the governing distribution of a system by analysis of a code-length game between the system (``Nature'') and the observer \cite{Topsoe_2002, Topsoe_2004}. For a multivariate system of {\it iid} random variables, which take discrete values, this yields the multinomial distribution at game-theoretic equilibrium \cite{Topsoe_2002}. As in case (c), this could then be imposed as the governing distribution at a lower level of description.

\section{\label{concl}Conclusions}
%
This study examines the MaxProb principle, in which a system is represented by its distribution of highest probability. This can be interpreted as a generalized method of probabilistic inference, which does not provide certainty in its predictions, yet is always consistent with the rules of probability theory. In contrast, inference using the Kullback-Leibler cross-entropy or Shannon entropy functions, in cases in which the governing distribution is not multinomial and/or does not satisfy the asymptotic limits, can lead to inconsistencies. The MaxProb principle also gives rise to generalized combinatorial definitions of cross-entropy and entropy, an extension of the idea given by Boltzmann 130 years ago. The cross-entropy or entropy can therefore be {\it defined} so that its extremization, subject to the constraints, gives the ``most probable'' (``MaxProb'') realization of the system. This provides a purely probabilistic basis for MaxEnt and MinXEnt, which is independent of any information-theoretic justification. 

This work examines the origins of the governing distribution $\mathbb{P}$, including by (a) frequentist-like models (e.g.\ ball-in-box or urn models); (b) symmetry models; (c) prior MinXEnt models; (d) Kapur-Kesavan inverse models; and (e) game theoretic models.  It is shown that the combinatorial definition and MaxProb are consistent with these different approaches, and the ``subjective Bayesian'' definition of probability, yet is more broadly based and offers greater utility than traditional MaxEnt / MinXEnt based on the Shannon and Kullback-Leibler functions.

\section*{Acknowledgments}

The author thanks the European Commission for support as a Marie Curie Incoming International Fellow; Marian Grend\'ar, David Blower, Tony Bell and Flemming Tops\o e for specific arguments (as detailed above) and stimulating discussions; and the organisers and participants of MaxEnt07.

\end{document}